\documentclass[sigconf]{acmart}

\copyrightyear{2021}
\acmYear{2021}
\setcopyright{rightsretained}
\acmConference[ESEM '21]{ACM / IEEE International Symposium on Empirical 
Software Engineering and Measurement (ESEM)}{October 11--15, 2021}{Bari, Italy}
\acmBooktitle{ACM / IEEE International Symposium on Empirical Software 
Engineering and Measurement (ESEM) (ESEM '21), October 11--15, 2021, Bari, 
Italy}\acmDOI{10.1145/3475716.3484195}
\acmISBN{978-1-4503-8665-4/21/10}

\input{meta/packages}

\newcommand{\footurl}[1]{{\footnote{\url{#1}}}}

\newcommand{\checknum}[1]{{\color{black}{#1}}}

\newcommand{\licma}{\textit{LICMA}}
\newcommand{\cryptorex}{\textit{CryptoREX}}
\newcommand{\cryptolint}{\textit{CryptoLint}}

\newcommand{\cognicryptsast}{\textit{CogniCrypt\textsubscript{SAST}}}
\newcommand{\cryptoguard}{\textit{Cryptoguard}}

\newcommand{\babelfish}{\textit{Babelfish}}

\newcommand{\support}{{\color{green!70!black}{\CIRCLE}}}
\newcommand{\nosupport}{{\color{red!70!black}{\Circle}}}
\newcommand{\secabstact}[1]{\textbf{#1:}}
\newcommand{\rone}{§1}
\newcommand{\rtwo}{§2}
\newcommand{\rthree}{§3}
\newcommand{\rfour}{§4}
\newcommand{\rfive}{§5}
\newcommand{\rsix}{§6}
\newcommand{\cam}[1]{{\color{black}#1}}

\newtcbtheorem{obs}{Observation}%
{colback=gray!5,colframe=gray!35!black,fonttitle=\bfseries}{th}

\begin{document}
\title{Python Crypto Misuses in the Wild}

\author{Anna-Katharina Wickert}
\orcid{ https://orcid.org/0000-0002-1441-2423}
\email{wickert@cs.tu-darmstadt.de}
\affiliation{%
    \institution{Technische Universität Darmstadt}
    \streetaddress{Hochschulstraße AB}
    \city{Darmstadt}
    \country{Germany}
}
\author{Lars Baumgärtner}
\orcid{ https://orcid.org/0000-0002-5805-2773}
\email{baumgaertner@cs.tu-darmstadt.de}
\affiliation{%
    \institution{Technische Universität Darmstadt}
    \streetaddress{Hochschulstraße AB}
    \city{Darmstadt}
    \country{Germany}
}
\author{Florian Breitfelder}
\orcid{https://orcid.org/0000-0003-2337-1819}
\email{florian.breitfelder@tu-darmstadt.de}
\affiliation{%
    \institution{Technische Universität Darmstadt}
    \streetaddress{Hochschulstraße AB}
    \city{Darmstadt}
    \country{Germany}
}
\author{Mira Mezini}
\orcid{https://orcid.org/0000-0001-6563-7537}
\email{mezini@cs.tu-darmstadt.de}
\affiliation{%
    \institution{Technische Universität Darmstadt}
    \streetaddress{Hochschulstraße AB}
    \city{Darmstadt}
    \country{Germany}
}

\renewcommand{\shortauthors}{Wickert et al.}

\begin{abstract}
\secabstact{Background}
Previous studies have shown that up to \checknum{99.59~\%} of the Java apps using crypto APIs misuse the API at least once. 
However, these studies have been conducted on Java and C, while empirical studies for other languages are missing. 
For example, a \cam{controlled} user study with crypto tasks in Python has shown that \checknum{68.5~\%} of the professional developers write a secure solution for a crypto task. 
\secabstact{Aims}
To understand if this observation holds for real-world code, we conducted a study of crypto misuses in Python. %
\secabstact{Method}
We developed a static analysis tool that covers common misuses of \checknum{5} different Python crypto APIs.
With this analysis, we analyzed \checknum{895} popular Python projects from GitHub and \checknum{51} MicroPython projects for embedded devices. 
Further, we compared our results with the findings of previous studies \cam{.}
\secabstact{Results}
Our analysis reveals that \checknum{52.26~\%} of the Python projects have at least one misuse\cam{. }%
Further, some Python crypto libraries' API design helps 
developers from misusing crypto functions, which were much more common in studies conducted with Java and C code. 
\secabstact{Conclusion}
We conclude that we can see a positive impact of the good API design on crypto misuses for Python applications.
Further, our analysis of MicroPython projects reveals the importance of hybrid analyses. 
\end{abstract}



\maketitle

\section{Introduction}
\label{sec:intro}

Crypto\cam{graphy, hereafter crypto,} is widely used nowadays to protect our data and ensure confidentiality. 
For example, without crypto, we would not be able to securely use online banking or do online shopping. 
Unfortunately, previous research results show that crypto is often used in an insecure way~\cite{chatzikonstantinou2016evaluation, Egele.2013, DBLP:journals/corr/abs-1810-09065, lazar2014does, nadi2016jumping}. 
One such problem is the choice of an insecure parameter, like an insecure block mode, for crypto primitives like encryption. 
Many static analysis tools exist to identify these misuses such as   \cryptorex{}~\cite{zhang2019cryptorex}, \cryptolint{}~\cite{Egele.2013}, \cognicryptsast{}~\cite{krger_et_al:LIPIcs:2018:9215}, and \cryptoguard{}~\cite{rahaman2019cryptoguard}.

While these tools and the respective in-the-wild studies concentrate on Java and C, user studies suggest that the existing Python APIs reduce the number of crypto misuses. 
Acar et al.~\cite{acar2017security} conducted an experiment with \checknum{307} GitHub users which had to solve \checknum{3} crypto-related development tasks. 
They observed that \checknum{68.5~\%} of the professional developers wrote a secure solution in Python for the given task. 
Within a controlled experiment with \checknum{256} Python developers that tried to solve simple crypto tasks, Acar et al.~\cite{acar2017comparing} identified that a simple API design, like the Python library \textit{cryptography}, supports developers in writing secure code. 
However, no empirical in-the-wild study has yet confirmed that crypto misuses in Python occur less frequently than in Java or C. 

To empirically evaluate crypto misuses in Python, we introduce \licma{}, \cam{a} multi-language analysis framework with support for \checknum{5} different Python crypto APIs and Java's JCA API. 
We provide \checknum{5} different rules~\cite{Egele.2013} for all Python APIs and \checknum{6} different rules~\cite{Egele.2013} for JCA to detect the most common crypto misuses.
With \licma{}, we analyzed \checknum{895} popular Python apps from GitHub and \checknum{51} MicroPython projects to gain insights into misuses in Python.
We identified that \checknum{52.26~\%} of the Python GitHub apps with crypto usages have at least one misuse causing \checknum{1,501} misuses.
In total, only \checknum{7~\%} of the misuses are within the application code itself, while the \checknum{remaining}  misuses are introduced by dependencies. 
Further, our study of MicroPython projects reveals that developers in the embedded domain tend to use crypto via C code.
Thus, revealing the importance of hybrid static analyses, which can track program information, e.g., a call graph, across multiple languages~\cite{mushtaq2017multilingual, ferrara2019cross}. 
\begin{table*}[!htbp]
    \caption{Six commonly discussed crypto misuses \cam{in Java and C}~\cite{Egele.2013,zhang2019cryptorex} with an example of a violation in Python. }
    \label{tab:back_rules}
    \centering
    \begin{tabular}{p{0.5cm} p{8cm} p{8cm} }
         \toprule
         ID & Rule & Python: Violation Example  \\
         \midrule
         §1 & Do not use electronic code book (ECB) mode for encryption. & \lstinline!aes = AES.new(key, AES.MODE_ECB)!  \\ 
         §2 & Do not use a non-random initiliazation vector (IV) for ciphertext block chaining (CBC) encryption. & \lstinline!aes = AES.new(key, AES.MODE_CBC, b'\0' * 16)! \\
         §3 & Do not use constant encryption keys. & \lstinline!aes = AES.new(b'\0' * 32, AES.MODE_CBC, iv)! \\
         §4 & Do not use constant salts for password-based encryption (PBE). & \lstinline!kdf = PBKDF2HMAC(hashes.SHA256(), 32, b'\0' * 32, 10000)!  \\
         §5 & Do not use fewer than 1,000 iterations for PBE. & \lstinline!kdf = PBKDF2HMAC(hashes.SHA256(), 32, salt, 1)!  \\
         §6 & Do not use static seeds to initialize secure random generator. & \noindent\cam{\small{Due to API design only possible in  Java~\cite{Egele.2013} and C/C++~\cite{zhang2019cryptorex}}}\\
         \bottomrule
    \end{tabular}
\end{table*}

\cam{To further improve our understanding whether Python APIs are less prone to crypto misuses, we make the following contributions:}
\begin{itemize}
\item A novel, multi-language analysis tool to detect crypto misuses in Python and Java. 
For Python we cover crypto misuses for \checknum{5} common Python crypto APIs and for Java the standard API JCA. 
\item An empirical study of crypto misuses in the \checknum{895} most popular Python applications on GitHub revealing
\checknum{1,501} misuses. 
\item A comparison of our findings in Python applications with previous studies about crypto misuses in-the-wild for Android Apps and firmware images in C. 
\cam{We observed that most Python applications are more secure and the distribution between the concrete types of misuses differ a lot.}
\item An empirical study of crypto misuses in MicroPython projects which reveals the importance of hybrid static analyses. 
\item A replication package including both data sets used for our study, the results of our analysis, and the code of \licma{}\footnote{ \url{dx.doi.org/10.6084/m9.figshare.16499085}}.  
\end{itemize}

\section{Background}
\label{sec:background}

\label{sec:bck-cryptolintRules}

Crypto libraries enable developers to use crypto primitives, like symmetric key encryption or password-based encryption (PBE), in their application. 
However, studies have shown that developers struggle to use those APIs correctly and securely.
For example, Krüger et al.~\cite{krger_et_al:LIPIcs:2018:9215} show that \checknum{95~\%} of the Android applications using crypto have at least one crypto misuse.
In the remainder of the paper, we define a crypto misuse, in short misuse, as a usage of a crypto library which is a formally correct use of the API, e.g., not causing a crash, but is \cam{unsafe}. %
In this paper, we focus on six commonly discussed crypto misuses~\cite{Egele.2013, zhang2019cryptorex} which we will introduce next. 
All rules including Python example code can be seen in Table~\ref{tab:back_rules}. 

For AES encryption, a block mode is used to combine several blocks for encryption and decryption.  
However, some block modes are considered insecure by experts. 
Rule 1 (\rone) and rule 2 (\rtwo) cover two insecure usages of block modes. 
\rone~prohibits the usage of \textit{ECB} which is deterministic and not secure.
\rtwo~focus on a non-random initialization vector (IV) for the block mode \textit{CBC}.
Due to the predictability of the IV, the usage of CBC is deterministic and not secure.
Further, a key protects confidential information.
Thus, rule 3 (\rthree) states that the key should be confidential and not a static one hard-coded in the source. 

For PBE, a salt and the number of iterations harden the password against dictionary attacks. 
Rule 4 (\rfour) highlights that a non-random salt reduces the security of the crypto primitive to using no salt, which is considered insecure. 
Rule 5 (\rfive) concentrates on the fact that several iterations are required to harden the crypto primitive against attacks. 
Thus, Egele et al.~\cite{Egele.2013} suggest the minimum recommended value of 1,000 iterations as defined in the PBE standard PKCS\#5.

A seed can ensure that the output of a random number generator is non-deterministic.
\cam{Rule 6 (\rsix) covers that static seeds lead to deterministic outputs.}
\cam{Common Java~\cite{Egele.2013} and C/C++~\cite{zhang2019cryptorex} libraries expect a seed, while the analyzed Python APIs avoid this by design.}

\section{Design and Implementation of \licma{}}
\label{sec:appr}

\cam{In this section, we describe the design of our static analysis tool \licma{}, and discuss the implementation in more detail.}

\begin{figure*}
	\centering
		\scalebox{0.90}{
			\smartdiagramset{
				set color list={white, white, white, white},
				uniform arrow color=true,
				arrow color=gray,
				back arrow disabled=true,
				module minimum width=3cm,
				circular distance=6cm,
				module x sep=4cm,
				text width=3cm,
				additions={
					additional item offset=1cm,
					additional item fill color=white,
					additional item border color=gray,
					additional arrow color=gray,
					additional item width=2cm,
					additional item height=1cm,
					additional item text width=2cm,
					additional item bottom color=gray,
					additional item shadow=drop shadow,
				}
		}
		\smartdiagramadd[flow diagram:horizontal]{\textbf{Input:} source file, Parsing with Babelfish, \licma{} analysis, \textbf{Output:} CSV file with misuses}{
			below of module2/Java analysis component, below of module3/Python analysis component, below of module4/\dots
		}
		\smartdiagramconnect{<-}{module3/additional-module1}
		\smartdiagramconnect{<-}{module3/additional-module2}
		\smartdiagramconnect{<-}{module3/additional-module3}
		}
	\vspace{20mm} %
	\caption{\licma{} analysis framework.}\label{fig:licma-analysis-steps}
\end{figure*}

\subsection{Design}
\label{sec:appr:design}

A general overview of \licma{} is given in Figure~\ref{fig:licma-analysis-steps}.
First, we parse a source code file into the respective Abstract Syntax Tree~(AST).
More specifically, we use \babelfish{}~\footnote{https://github.com/bblfsh/bblfshd} to create a Universal Abstract Syntax Tree (UAST) which combines language-independent AST elements with language-specific elements. 
For simplicity, we use the term AST in the remainder of the paper. 
Second, we apply the \licma{} analysis upon the AST to identify potential misuses of our crypto rules.

Based upon the rule defining a misuse, the analysis checks for a violation of the rule within the source code and triggers a backward analysis for this task.  
The backward slice is created by filtering the AST with the help of XPath\footnote{\url{https://www.w3.org/TR/xpath/}} queries%
, and works as follows:  
First, the backward slicing algorithm (BSA) identifies all source code lines that are referred within the respective rule. 
An example of the slicing criterion is a function call parameter like the key for a crypto function. 
Second, the BSA determines for all function calls if the parameter is either hard-coded, a local assignment, or a global assignment. 
If one of the three cases is fulfilled, the corresponding value is returned. 
This value is checked against a function defined in the rule, e.g., if the value is smaller than \checknum{1,000} for \rfive{}.
In the negative case, the BSA looks for the caller of the function, and checks the caller's parameters as described above. 
The algorithm stops if a value is returned or no further callers to analyse are available, and returns the result of the analysis. 
For its reports, \licma{} distinguishes between a potential misuse if it can not resolve the value of interest due to missing callers, and a definite misuse if it is resolved to an insecure option. 

\subsection{Implementation}
\label{sec:appr:impl}

For our study, we implemented Python and Java analysis components. 
For Python, we cover \checknum{5} different crypto modules: \textit{cryptography, M2Crypto, PyCrypto, PyNaCl, ucryptolib}.
This selection is based upon the inspected Python modules by Acar et al.~\cite{acar2017comparing} which is based on the libraries' popularity, their possibility to solve common crypto tasks, and a mix of usability focus, e.g., API-design with ignoring usability up to usability as a main focus. 
However, we dropped the deprecated module \textit{Keyczar}, and added the MicroPython library \textit{ucryptolib}.

As the six rules are defined with the \textit{JCA} in mind~\cite{Egele.2013}, we could not implement all of these rules for Python. 
For none of the Python modules, a misuse of \rsix{} is possible due to the design of the APIs 
as there are no secure random number generators present that can be initialized with a static seed.

Further, \licma{} supports only \rthree{}~and \rfour{}~for \textit{PyNaCl} and only \rone, \rtwo~and \rthree~for \textit{ucryptolib}. 
The derivation is due to a different API design which avoids the respective misuses. 
We present an overview of the implemented APIs and the covered rules in Table~\ref{tab:rules-to-lib}.

\cam{Our implementation of \licma{} is available on GitHub\footnote{\url{https://github.com/stg-tud/licma}}.}

\begin{table}[t]
    \caption{An overview of the libraries supported in \licma{} and the rules of potential misuses.}
    \label{tab:rules-to-lib}
    \centering
    \begin{tabular}{lllllll}
    \toprule
    Library & §1 &  §2 & §3 & §4 & §5 & §6 \\
    \midrule
    JCA & \support{} & \support{} & \support{} & \support{} & \support{} & \support{}\\ 
    \arrayrulecolor{black!40}\midrule
    cryptography & \support{} & \support{} & \support{} & \support{} & \support{} & \nosupport{}\\
    M2Crypto & \support{} & \support{} & \support{} & \support{} & \support{} & \nosupport{}\\
    PyCrypto & \support{} & \support{} & \support{} & \support{} & \support{} & \nosupport{} \\
    PyNaCI & \nosupport{} & \nosupport{} & \support{} & \support{} & \nosupport{} & \nosupport{} \\
    ucryptolib & \support{} & \support{} & \support{} & \nosupport{} & \nosupport{} & \nosupport{} \\
    \bottomrule     
    \end{tabular}
    \vspace{2mm}
    {\\ \small{\support{} indicated that the library can cover misuses of the rule, while \\
    \nosupport{} indicates that this misuse is not possible through the API.}}
\end{table}
\section{Methodology}

To analyze Python applications, we constructed two distinct data sets of popular Python and MicroPython projects. 
Furthermore, we compared our findings in Python programs with previous studies about Java and C code. 

\subsection{Searching and Downloading Python Apps}

Both data sets represent very different domains where Python is used, ranging from server and desktop use to low-level embedded code.
How we selected the projects in both data sets for our empirical study is described in the following.	

\subsubsection{Python Projects from GitHub}

For our evaluation of crypto misuses in Python code we focus on open-source code. 
Thus, we crawled and downloaded the top \checknum{895} Python repositories from GitHub sorted by stars. 
To further understand the influence of dependencies, we downloaded them with Pythons standard dependency manager \textit{pip} for each project. %
Afterwards, we ended up with \checknum{14,442} Python packages of which \checknum{3,420} are unique.

As our analysis works upon a per-file basis, we reduced our set to only those source code files that include the function calls referenced in our rules, e.g., \lstinline[language=C]!AES.new(...)!. %
In addition, we filter for production code and ignore test code which should be non-existent during the execution of the application.
After applying these \checknum{2} filter steps, we ended up with \checknum{946} source files from \checknum{155} different repositories. 
Unfortunately, \babelfish{} was unable to parse \checknum{35} of these files, and reached the maximum recursion depth for the AST XPath queries for at least one rule in \checknum{50} files. 
These \checknum{85} parsing failures are distributed amongst \checknum{61} different projects.
However, for each of the projects at least \checknum{1} file with a crypto usage was analyzed successfully. 
In total, we successfully analyzed \checknum{861} different files within \checknum{155} Python repositories with \licma{}.

\subsubsection{Curated Top MicroPython Projects}
\label{sec:meth:micro}

As an extension to our Python application set, we crawled \checknum{51} MicroPython projects which are stated as the top announced MicroPython projects\footnote{\url{https://awesomeopensource.com/projects/micropython}}. %
Like for the regular Python applications, we downloaded all dependencies with \textit{pip} and got \checknum{113} dependencies with \checknum{1} duplicate dependency.
Afterwards, we applied the same filter steps as before: 
The usage of crypto and the exclusion of test files. 
These steps, resulted in \checknum{5} files which seem to use the Python crypto libraries supported by \licma. 
Note that we included the MicroPython crypto library \textit{ucryptolib} in \licma{} and our filtering steps. 
To further understand this small number of potential usages, we also analyzed our data set of MicroPython applications manually. 
This analysis reveals that we potentially missed \checknum{five} crypto usages.

\subsection{Comparison with Previous Studies}

To understand the differences between crypto misuses \cam{for \rone~to \rsix, cf. Table~\ref{tab:back_rules},} \cam{in Python} and previous studies in Java, with the analysis \cryptolint,~\cite{Egele.2013} and C/C++, with the analysis \cryptorex,~\cite{zhang2019cryptorex}, we \cam{compared the reported results}. %
As we concentrated on the same rule set, we only need a few adjustments to compare the results.
First, for our meta-analysis we exclude \rsix~since the \checknum{5} analyzed Python modules avoid this misuse by design. 
Second, we merge the results for \rone~of Egele et al.~\cite{Egele.2013} as they split their result into two different cases: 
The explicit use of the block mode ECB on one side and the implicit use of this block mode due to the API design on the other. 
Third, due to the design of our analysis, we only consider definite findings.
\cam{\cryptolint~and \cryptorex~do not distinguish between potential misuses and definite ones.}
Fourth, to enable a fair comparison, we compare only percentages rather than absolute numbers, as we are interested in the general distribution and the influence of %
API design on crypto misuses. 
We choose to compare the studies on the percentage of applications using crypto and having at least one misuse of a respective rule as introduced by Egele et al.~\cite{Egele.2013}.
Unfortunately, Zhang et al.~\cite{zhang2019cryptorex} only reports details for the successfully unpacked firmware images before filtering for crypto usages. 
\section{Evaluation}
\label{sec:eval}

In this Section, we present our evaluation of crypto misuses in real-world Python applications 
from GitHub and 
MicroPython projects. 
	
\subsection{GitHub Python Projects}
\label{sec:eval:python}

Overall, \licma{} identified \checknum{1,501} possible misuses in our data set of Python applications  
attributed to \checknum{81} repositories. 
Thus, \checknum{52.26~\%} of the \checknum{155} analyzed Python applications that contain crypto usages have at least one misuse. 

As discussed in Section~\ref{sec:appr:design}, we distinguish between potential and definite misuses. 
While a potential misuse requires a manual inspection to decide whether it is harmful, 
a definite misuse indicates that the analysis was able to resolve the respective crypto parameter. 
Thus, we know that a rule of \licma{} is definitely violated by the respective API call. 
We identified \checknum{85} definite misuses which could be identified within one class file and thus are local. 
The remaining \checknum{1,416} misuses are potential misuses. 

\begin{figure}[t]
	\centering
    \includegraphics[width=0.93\columnwidth]{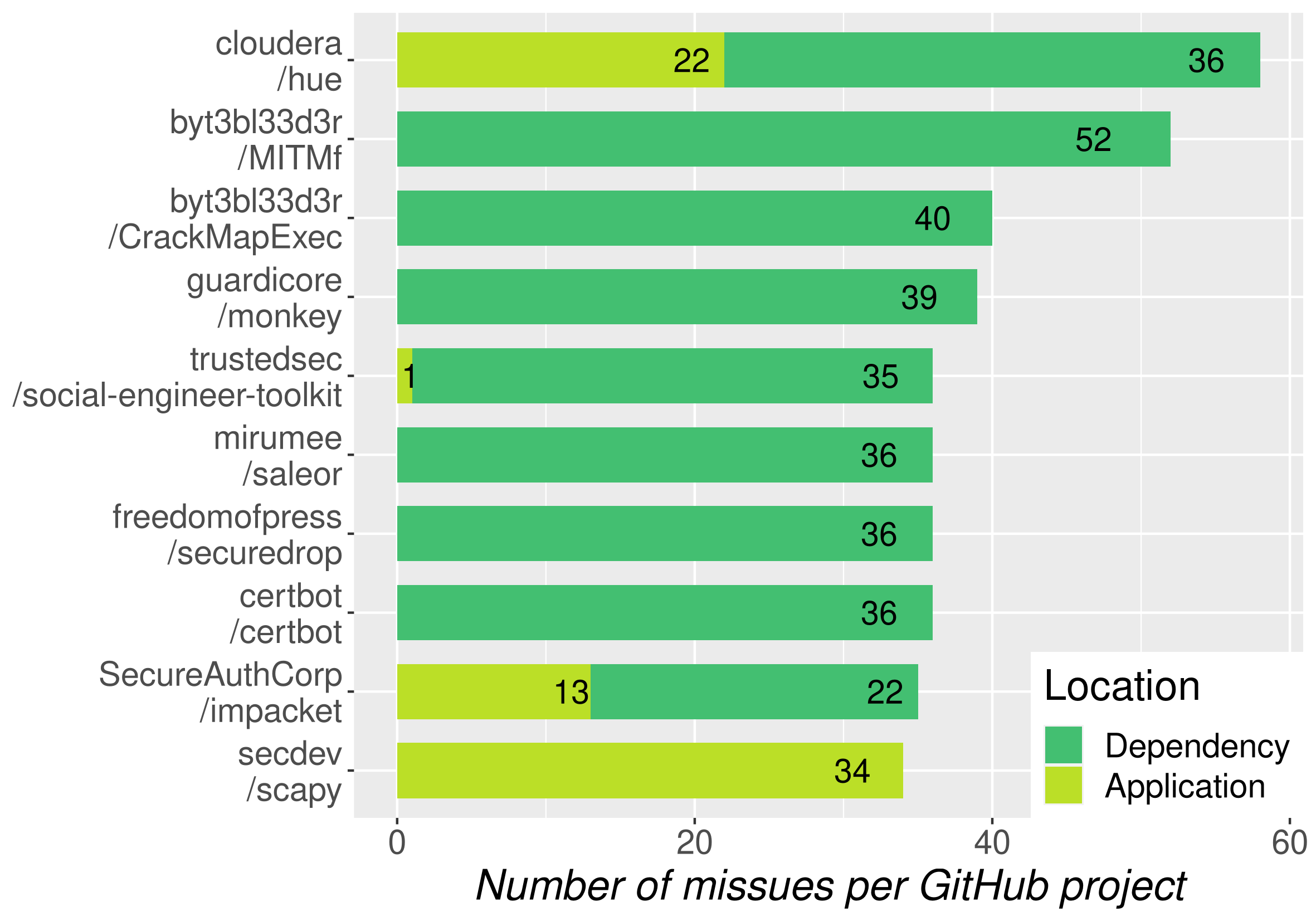}
	\caption{Python projects with 30 or more misuses.}
	\label{fig:repositories-with-misuses}
\end{figure}

\begin{obs}{}{}
In total, \checknum{52.26~\%} of the Python projects using crypto APIs contain at least a potential misuse. 
Only \checknum{5.66~\%} of the misuses are local (within one class file). 
\end{obs}

\subsubsection{Dependencies}

From the \checknum{1,501} misuses, only \checknum{7.00~\%} are within the application code and not in dependencies. 
These misuses are within \checknum{14.81~\%} of the applications with at least one misuse. 
The remaining misuses are found in dependencies and can be reduced to \checknum{290} unique misuses. 
Thus, developers introduce most of their misuses by using dependencies rather than using the respective crypto library directly.
In total, only \checknum{12} projects are affected by misuses in the application code itself.

To understand the influence of dependencies within the applications with the most misuses, we inspected the \checknum{10} Python projects with more than \checknum{30} misuses. 
Figure~\ref{fig:repositories-with-misuses} confirms the previous observation that most of the misuses are in dependencies and only a few projects use a crypto library directly. 
The \textit{Scapy} repository\footnote{ \url{https://github.com/secdev/scapy}} is an exception as all misuses are in its code. 
Our investigation reveals that this repository is often used as a dependency by other projects.
Thus, these findings can be attributed to dependencies as well.

While, the previous results focused on the projects, we also inspected the dependencies causing most of the misuses. 
In total, \checknum{5} of the observed dependencies are responsible for a misuse in more than \checknum{10} different projects. 
For \checknum{34} projects we observe a misuse within the repository \textit{Scapy} which was the only analyzed repository in Figure~\ref{fig:repositories-with-misuses} without any misuse in its dependencies. 
Thus, confirming our previous observation about this project. 

\begin{obs}{}{}
Only \checknum{14.81~\%} of the projects directly contain a misuse of a crypto API. 
The rest is introduced through third-party code.
Thus, it is important for developers and analysts to understand the security implications of the libraries used. 
\end{obs}
\vspace{-1mm}

\subsubsection{Rules and Python Cryptographic Libraries}

\begin{figure}[t!]
	\centering
    \includegraphics[width=0.95\columnwidth]{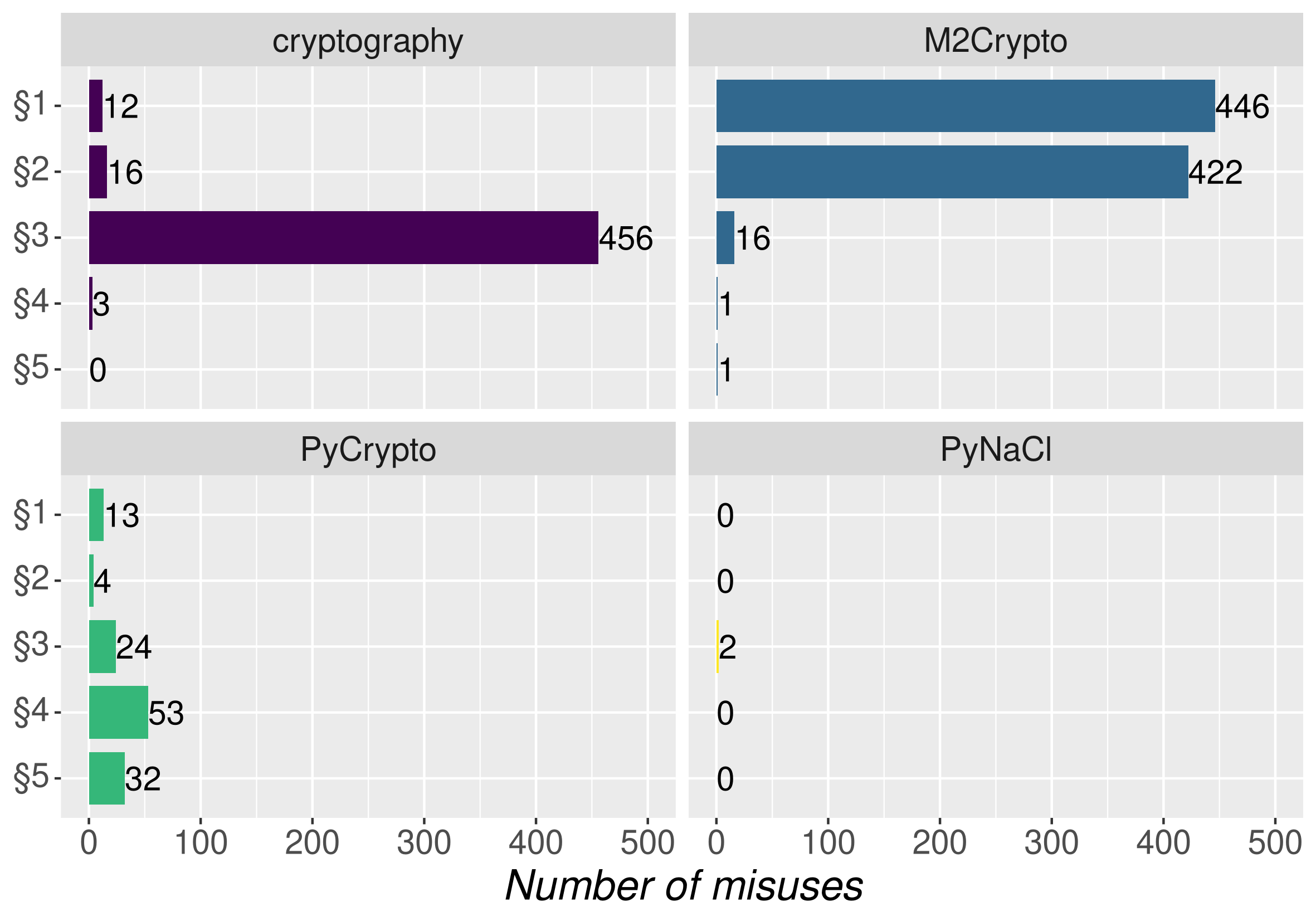}
	\caption{The number of misuses found per rule for the Python libraries \textit{cryptography}, \textit{M2Crypto}, \textit{PyCrypto}, \textit{PyNaCl}. The later library avoids per design misuses for \rone{}, \rtwo{} and \rfive{}.}\label{fig:licma-rules-distribution}
\end{figure}

In order to get a better understanding of the underlying reasons of the misuses, we evaluated how often a misuse per rule and library occurs~(Fig.~\ref{fig:licma-rules-distribution}). 
Our analysis reveals that most of the misuses are related to the use of different block modes, \rone~and \rtwo, of the \textit{M2Crypto} library, and constant encryption keys, \rthree, for the \textit{cryptography} library. 
We assume that the few numbers of misuses of \rone~and \rtwo~of \textit{cryptography} are due to the design of the library.
The library suggests to use a high-level symmetric encryption class, called \textit{Fernet}, instead of the low-level symmetric encryption classes which would enable the respective misuses. 
Most of the misuses due to insecure PBE configurations, \rfour~and \rfive, are by developers using the library \textit{PyCrypto}. 
While, none of the \checknum{3} previously mentioned libraries make it impossible to produce a misuse of one of our \checknum{5} rules, the library \checknum{PyNaCl} completely prevents misuses for \rone, \rtwo, and \rfive. 
In our study, we found only \checknum{2} instances of a misuse due to a constant encryption key for \textit{PyNaCl}. 

\begin{obs}{}{}
The design of the Python crypto API \textit{cryptography} supports developers in avoiding misuses due to an insecure block mode for AES encryption. 
\end{obs}

\subsubsection{Reasons for Definite Misuses}

Among the definite misuses, there is at least one misuse for \checknum{all} previously discussed rules. 
We identified \checknum{13} definite misuses in \checknum{5} different projects which use the ECB encryption mode (\rone).
In all cases, the mode is passed explicitly with the parameter and not implicitly as in Java~\cite{Egele.2013}.
For \checknum{8} misuses we observed that a static IV is used, e.g. zero bytes, and thus resulting in an insecure encryption with the block mode CBC (\rtwo). 
Furthermore, we identified that the \textit{scapy} project which is also commonly used as a dependency uses a constant encryption key resulting in \checknum{14} misuses (\rthree). 
For example, we found a zero byte-array as key. 

For password-based encryption, we identified \checknum{18} misuses within \checknum{14} projects which pass a static salt instead of a randomly generated one (\rfour). 
In total, we identify \checknum{32} misuses which are due to requesting only \checknum{1} iteration instead of an value greater than \checknum{1,000} as recommended (\rfive). 
Thus, the process of generating a password is faster but very insecure, e.g., due to dictionary attacks. 

\begin{obs}{}{}
\cam{As a result of  \checknum{58.82~\%} of the definite misuses, passwords are vulnerable to dictionary attacks (\rfour~and~\rfive)}.%
\end{obs}
	
\subsection{MicroPython}

When we applied \licma{} upon the \checknum{5} source files containing crypto API usages of the MicroPython data set, we identified \checknum{no} misuse. 
For this reason, we inspected the MicroPython repositories for usages of other crypto functions not covered by \licma{}~and identified \checknum{5} additional files. 
We notice that the crypto module \texttt{ucryptolib} which is provided by MicroPython, is only used by tests in \checknum{2} projects. 
For the remaining \checknum{3} findings, the crypto functions are written in C rather than Python.
Thus, these files were removed due to our filter steps described in Section~\ref{sec:meth:micro}.

Our small analysis of MicroPython projects shows that for a further exploration of MicroPython applications we need to consider custom implementations of AES in Python and C. 
This seems to be a common pattern for embedded code where performance is important and low-level code is often shipped as custom C blobs.
Thus, we can observe the importance of hybrid analyses approaches~\cite{ferrara2019cross, mushtaq2017multilingual}. 

\section{Comparison with previous Studies}

\begin{figure}[t!]
	\centering
   \includegraphics[width=0.95\columnwidth]{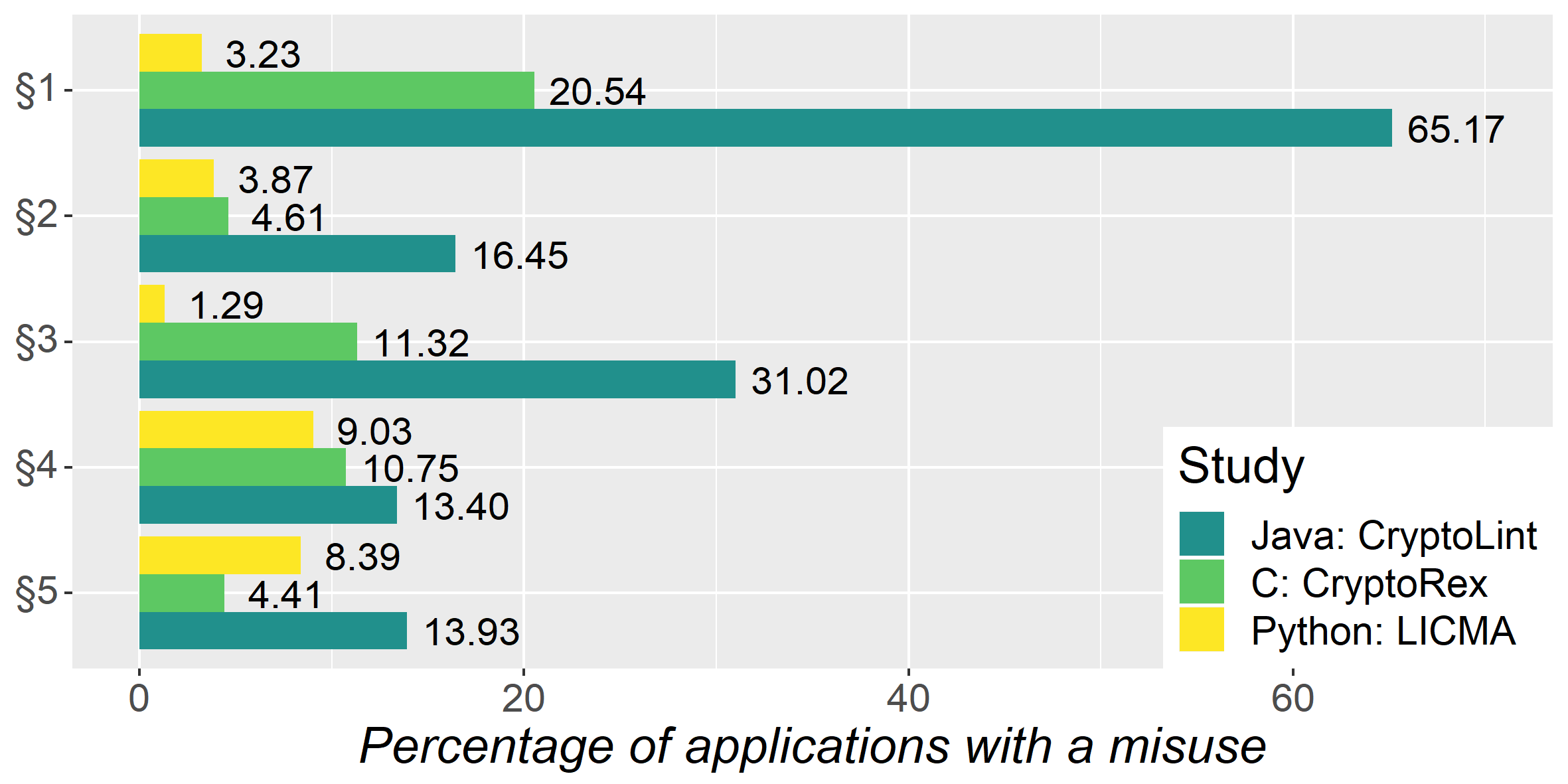} 
		\caption{Comparison of our evaluation results with the results of \cryptolint{}~\cite{Egele.2013} and \cryptorex{}~\cite{zhang2019cryptorex}. 
		}\label{fig:meta}
\end{figure}
	
As one motivation of this paper was to empirically shed light on the question whether Python crypto libraries
help developers in writing more secure code than previous empirical studies in Java or C have revealed, we compare our results to the findings of these studies. 
For all applications which use crypto, we observe more secure applications than reported by Egele et al.~\cite{Egele.2013}.
While for Android apps using crypto libraries at least one misuse occurs in \checknum{87.90~\%} of the apps, we observe "only" \checknum{52.26~\%} of Python applications with a misuse. 
Unfortunately, Zhang et al.~\cite{zhang2019cryptorex} only report the number of C firmware images they started to analyze and not explicitly mention how many of these actually use a crypto API. 
Thus, we only know that \checknum{24.18~\%} of the analyzed firmware images have at least one misuse. 

In Figure~\ref{fig:meta} we present the percentage of applications per study and per rule with a misuse.  
In general, we observed that misuses in Python occur less frequently than for Java and C. 
The misuse of \textit{ECB} as an encryption mode is the most-misused in Java and C, and is significantly less with \checknum{3.23~\%} of the Python applications. 
We hypothesize that this difference is due to the design of the libraries as discussed in Section~\ref{sec:eval:python}. 

\begin{obs}{}{}
Our comparison reveals that the design of the crypto APIs in Python helps developers to avoid common misuses on the use of block modes for AES encryption in their code. 
\end{obs}

\section{Threats to Validity}
\label{sec:threatsToValidity}

We evaluated top GitHub Python projects and it may be that our results fail to generalize on specialized Python applications\cam{.}%
For our data set on MicroPython applications, we also concentrated on popular projects. 
Thus, our insights may not generalize to less popular or closed-source projects.  
However, we believe that our results provide first interesting insights on crypto misuses in Python. 

Currently, our analysis is limited to capabilities of \babelfish, especially the recursive maximum depth of its filter function. 
Furthermore, currently \babelfish{} only creates an AST for a single file. %
Thus, our analysis fails to resolve misuse over multiple files.  
We hope that these limitations can be lifted through further development of \babelfish{}. 
\cam{These improvements will hopefully help to reduce the number of false-positives in the potential misuses.}
\cam{Furthermore, it may be that our static analysis missed some misuses as Python is a dynamic typed language.} 

\cam{We compare} different application types of studies conducted in different years. 
Thus, it may be that the results might change when conducted on the same kind of applications now. 
\cam{Further, the results may differ due to the effect of different application domains and different analysis frameworks.}
Moreover, the percentages of applications with at least one misuse per rule that we used from Zhang et al.~\cite{zhang2019cryptorex} might be too positive for C, as the number of firmware images with crypto usages is not explicitly reported.

\section{Related Work}
\label{sec:related-work}

Several, previous studies show that crypto misuses occur frequently in different languages and platforms. 
Egele et al.~\cite{Egele.2013}, Krüger et al.~\cite{krger_et_al:LIPIcs:2018:9215}, Rahaman et al.~\cite{rahaman2019cryptoguard}, and Hazhirpasand et al.~\cite{hazhirpasand2019Java} analyzed Java and Android applications.
They reported that \checknum{84.78~\%} up to \checknum{99.59~\%} of the applications using crypto have at least one misuse. 
Zhang et al.~\cite{zhang2019cryptorex} analyzed Internet of Things (IoT) device firmwares written in C/C++, from which \checknum{24.2~\%} contain at least one misuse. 

Previous work introducing new crypto misuse analyses either improve static analysis approaches for crypto misuse detection or introduce these to new languages imposing new challenges. 
\cryptolint{}~\cite{Egele.2013} is the first (closed-source) static analysis for crypto misuses for Android applications introducing the six rules for crypto misuses, \cam{c.f. Table~\ref{tab:back_rules}.} %
While this analysis is built upon a deny-listing approach, \cognicryptsast{}~\cite{krger_et_al:LIPIcs:2018:9215} introduces an allow-listing approach covering the standard Java library\cam{, BouncyCastle and Tink} to analyze Java and Android applications for 
misuses. 
The focus of the
analysis \cryptoguard{}~\cite{rahaman2019cryptoguard} is a scalable deny-listing Java analysis for crypto misuses extending the 
\cam{rules}
implemented in \cryptolint{}~\cite{Egele.2013}. 
\cryptorex{} is a framework for firmware written in C/C++ which covers the rules introduces by \cryptolint{}~\cite{Egele.2013}.

Acar et al.~\cite{acar2017comparing} conducted a user study with \checknum{5} different Python crypto APIs to analyze how developers perform on \checknum{5} crypto tasks with a pre-selected API.
Their study reveals that APIs with a usability focus for security result in significant more secure code\cam{.}
In a similar study, Acar et al.~\cite{acar2017security} analyzed the security of \checknum{3} different crypto tasks and identified that more usable libraries resulted only in insecure solutions for encryption in \checknum{12.7~\%} of the cases.

\section{Conclusion}
\label{sec:concl}

In this paper, we presented the first empirical study of crypto misuses in Python. 
To conduct the study, we implemented the first multi-language analysis tool for crypto misuses with rules to detect common misuses of \checknum{five} different Python libraries as well as the standard Java library. 
We analyzed \checknum{895} popular Python apps from GitHub and \checknum{51} MicroPython projects to identify misuses.  
Our analysis revealed that \checknum{52.26~\%} of the projects using a crypto API, misuse the respective library. 
Furthermore, we observed that only \checknum{7~\%} of the \checknum{1,501} misuses are within the application code. 
The analysis of embedded applications written in MicroPython revealed the importance of hybrid analysis as the only crypto calls were in C code that got shipped with the projects.  

To get an impression on the differences between the different domains and languages analyzed in previous studies, %
\cam{w}e compared our results against the misuses reported for Android apps
~\cite{Egele.2013} and C firmware images~\cite{zhang2019cryptorex}. %
Our comparison confirms the impression that an \cam{opinionated API} design actually helps developers avoiding 
misuses. 

While we concentrated on the impact of a user-friendly API design for Python, future work can verify if these results generalize to other languages, like Rust and Go. 
Thus, extending \licma{} with new languages.
Further, it may be interesting to extend the currently implemented rules in \licma{} by an in-depth analysis of misuses of Python crypto APIs. 

\section*{Acknowledgments}

\cam{This research work has been co-funded by the Deutsche Forschungsgemeinschaft (DFG) – SFB 1119 CROSSING (236615297) and SFB 1053 MAKI (210487104), by the German Federal Ministry of Education and Research and the Hessen State Ministry for Higher Education, Research and the Arts within their joint support of the National Research Center for Applied Cybersecurity ATHENE, by the LOEWE
initiative (Hesse, Germany)
within the emergenCITY
center.}

\bibliographystyle{ACM-Reference-Format}
\bibliography{references, references-withoutZotero}


\begin{thebibliography}{13}


\ifx \showCODEN    \undefined \def \showCODEN     #1{\unskip}     \fi
\ifx \showDOI      \undefined \def \showDOI       #1{#1}\fi
\ifx \showISBNx    \undefined \def \showISBNx     #1{\unskip}     \fi
\ifx \showISBNxiii \undefined \def \showISBNxiii  #1{\unskip}     \fi
\ifx \showISSN     \undefined \def \showISSN      #1{\unskip}     \fi
\ifx \showLCCN     \undefined \def \showLCCN      #1{\unskip}     \fi
\ifx \shownote     \undefined \def \shownote      #1{#1}          \fi
\ifx \showarticletitle \undefined \def \showarticletitle #1{#1}   \fi
\ifx \showURL      \undefined \def \showURL       {\relax}        \fi
\providecommand\bibfield[2]{#2}
\providecommand\bibinfo[2]{#2}
\providecommand\natexlab[1]{#1}
\providecommand\showeprint[2][]{arXiv:#2}

\bibitem[\protect\citeauthoryear{Acar, Backes, Fahl, Garfinkel, Kim, Mazurek,
  and Stransky}{Acar et~al\mbox{.}}{2017a}]%
        {acar2017comparing}
\bibfield{author}{\bibinfo{person}{Yasemin Acar}, \bibinfo{person}{Michael
  Backes}, \bibinfo{person}{Sascha Fahl}, \bibinfo{person}{Simson Garfinkel},
  \bibinfo{person}{Doowon Kim}, \bibinfo{person}{Michelle~L Mazurek}, {and}
  \bibinfo{person}{Christian Stransky}.} \bibinfo{year}{2017}\natexlab{a}.
\newblock \showarticletitle{Comparing the usability of cryptographic apis}. In
  \bibinfo{booktitle}{\emph{IEEE Symposium on Security and Privacy (SP)}}.
  IEEE, \bibinfo{pages}{154--171}.
\newblock


\bibitem[\protect\citeauthoryear{Acar, Stransky, Wermke, Mazurek, and
  Fahl}{Acar et~al\mbox{.}}{2017b}]%
        {acar2017security}
\bibfield{author}{\bibinfo{person}{Yasemin Acar}, \bibinfo{person}{Christian
  Stransky}, \bibinfo{person}{Dominik Wermke}, \bibinfo{person}{Michelle~L
  Mazurek}, {and} \bibinfo{person}{Sascha Fahl}.}
  \bibinfo{year}{2017}\natexlab{b}.
\newblock \showarticletitle{Security developer studies with github users:
  Exploring a convenience sample}. In \bibinfo{booktitle}{\emph{Symposium on
  Usable Privacy and Security (SOUPS)}}. \bibinfo{pages}{81--95}.
\newblock


\bibitem[\protect\citeauthoryear{Chatzikonstantinou, Ntantogian, Karopoulos,
  and Xenakis}{Chatzikonstantinou et~al\mbox{.}}{2016}]%
        {chatzikonstantinou2016evaluation}
\bibfield{author}{\bibinfo{person}{Alexia Chatzikonstantinou},
  \bibinfo{person}{Christoforos Ntantogian}, \bibinfo{person}{Georgios
  Karopoulos}, {and} \bibinfo{person}{Christos Xenakis}.}
  \bibinfo{year}{2016}\natexlab{}.
\newblock \showarticletitle{Evaluation of cryptography usage in android
  applications}. In \bibinfo{booktitle}{\emph{EAI International Conference on
  Bio-inspired Information and Communications Technologies (formerly
  BIONETICS)}}. \bibinfo{pages}{83--90}.
\newblock


\bibitem[\protect\citeauthoryear{Egele, Brumley, Fratantonio, and
  Kruegel}{Egele et~al\mbox{.}}{2013}]%
        {Egele.2013}
\bibfield{author}{\bibinfo{person}{Manuel Egele}, \bibinfo{person}{David
  Brumley}, \bibinfo{person}{Yanick Fratantonio}, {and}
  \bibinfo{person}{Christopher Kruegel}.} \bibinfo{year}{2013}\natexlab{}.
\newblock \showarticletitle{An empirical study of cryptographic misuse in
  android applications}. In \bibinfo{booktitle}{\emph{ACM SIGSAC conference on
  Computer {\&} communications security (CCS)}},
  \bibfield{editor}{\bibinfo{person}{Ahmad-Reza Sadeghi},
  \bibinfo{person}{Virgil Gligor}, {and} \bibinfo{person}{Moti Yung}} (Eds.).
  \bibinfo{publisher}{ACM}, \bibinfo{pages}{73--84}.
\newblock
\showISBNx{9781450324779}


\bibitem[\protect\citeauthoryear{Ferrara, Mandal, Cortesi, and Spoto}{Ferrara
  et~al\mbox{.}}{2019}]%
        {ferrara2019cross}
\bibfield{author}{\bibinfo{person}{Pietro Ferrara}, \bibinfo{person}{Amit~Kr
  Mandal}, \bibinfo{person}{Agostino Cortesi}, {and} \bibinfo{person}{Fausto
  Spoto}.} \bibinfo{year}{2019}\natexlab{}.
\newblock \showarticletitle{Cross-programming language taint analysis for the
  iot ecosystem}.
\newblock \bibinfo{journal}{\emph{Electronic Communications of the EASST}}
  \bibinfo{volume}{77} (\bibinfo{year}{2019}).
\newblock


\bibitem[\protect\citeauthoryear{Hazhirpasand, Ghafari, and
  Nierstrasz}{Hazhirpasand et~al\mbox{.}}{2020}]%
        {hazhirpasand2019Java}
\bibfield{author}{\bibinfo{person}{Mohammadreza Hazhirpasand},
  \bibinfo{person}{Mohammad Ghafari}, {and} \bibinfo{person}{Oscar
  Nierstrasz}.} \bibinfo{year}{2020}\natexlab{}.
\newblock \showarticletitle{Java Cryptography Uses in the Wild}. In
  \bibinfo{booktitle}{\emph{ACM / IEEE International Symposium on Empirical
  Software Engineering and Measurement (ESEM)}}. \bibinfo{publisher}{ACM},
  Article \bibinfo{articleno}{40}.
\newblock
\showISBNx{9781450375801}


\bibitem[\protect\citeauthoryear{Kane, Lin, Chand, and Liu}{Kane
  et~al\mbox{.}}{2018}]%
        {DBLP:journals/corr/abs-1810-09065}
\bibfield{author}{\bibinfo{person}{Christopher Kane}, \bibinfo{person}{Bo Lin},
  \bibinfo{person}{Saksham Chand}, {and} \bibinfo{person}{Yanhong~A. Liu}.}
  \bibinfo{year}{2018}\natexlab{}.
\newblock \showarticletitle{High-level Cryptographic Abstractions}.
\newblock \bibinfo{journal}{\emph{CoRR}}  \bibinfo{volume}{abs/1810.09065}
  (\bibinfo{year}{2018}).
\newblock
\showeprint[arxiv]{1810.09065}
\urldef\tempurl%
\url{http://arxiv.org/abs/1810.09065}
\showURL{%
\tempurl}


\bibitem[\protect\citeauthoryear{Kr{\"u}ger, Sp{\"a}th, Ali, Bodden, and
  Mezini}{Kr{\"u}ger et~al\mbox{.}}{2018}]%
        {krger_et_al:LIPIcs:2018:9215}
\bibfield{author}{\bibinfo{person}{Stefan Kr{\"u}ger},
  \bibinfo{person}{Johannes Sp{\"a}th}, \bibinfo{person}{Karim Ali},
  \bibinfo{person}{Eric Bodden}, {and} \bibinfo{person}{Mira Mezini}.}
  \bibinfo{year}{2018}\natexlab{}.
\newblock \showarticletitle{{CrySL: An Extensible Approach to Validating the
  Correct Usage of Cryptographic APIs}}. In \bibinfo{booktitle}{\emph{European
  Conference on Object-Oriented Programming (ECOOP)}}
  \emph{(\bibinfo{series}{Leibniz International Proceedings in Informatics
  (LIPIcs)})}, Vol.~\bibinfo{volume}{109}. \bibinfo{publisher}{Schloss
  Dagstuhl--Leibniz-Zentrum fuer Informatik}, \bibinfo{address}{Dagstuhl,
  Germany}, \bibinfo{pages}{10:1--10:27}.
\newblock
\showISBNx{978-3-95977-079-8}
\showISSN{1868-8969}


\bibitem[\protect\citeauthoryear{Lazar, Chen, Wang, and Zeldovich}{Lazar
  et~al\mbox{.}}{2014}]%
        {lazar2014does}
\bibfield{author}{\bibinfo{person}{David Lazar}, \bibinfo{person}{Haogang
  Chen}, \bibinfo{person}{Xi Wang}, {and} \bibinfo{person}{Nickolai
  Zeldovich}.} \bibinfo{year}{2014}\natexlab{}.
\newblock \showarticletitle{Why does cryptographic software fail? A case study
  and open problems}. In \bibinfo{booktitle}{\emph{Asia-Pacific Workshop on
  Systems}}. \bibinfo{pages}{1--7}.
\newblock


\bibitem[\protect\citeauthoryear{Mushtaq, Rasool, and Shehzad}{Mushtaq
  et~al\mbox{.}}{2017}]%
        {mushtaq2017multilingual}
\bibfield{author}{\bibinfo{person}{Zaigham Mushtaq}, \bibinfo{person}{Ghulam
  Rasool}, {and} \bibinfo{person}{Balawal Shehzad}.}
  \bibinfo{year}{2017}\natexlab{}.
\newblock \showarticletitle{Multilingual source code analysis: A systematic
  literature review}.
\newblock \bibinfo{journal}{\emph{IEEE Access}}  \bibinfo{volume}{5}
  (\bibinfo{year}{2017}), \bibinfo{pages}{11307--11336}.
\newblock


\bibitem[\protect\citeauthoryear{Nadi, Kr{\"u}ger, Mezini, and Bodden}{Nadi
  et~al\mbox{.}}{2016}]%
        {nadi2016jumping}
\bibfield{author}{\bibinfo{person}{Sarah Nadi}, \bibinfo{person}{Stefan
  Kr{\"u}ger}, \bibinfo{person}{Mira Mezini}, {and} \bibinfo{person}{Eric
  Bodden}.} \bibinfo{year}{2016}\natexlab{}.
\newblock \showarticletitle{Jumping through hoops: Why do Java developers
  struggle with cryptography APIs?}. In \bibinfo{booktitle}{\emph{International
  Conference on Software Engineering (ICSE)}}. \bibinfo{pages}{935--946}.
\newblock


\bibitem[\protect\citeauthoryear{Rahaman, Xiao, Afrose, Shaon, Tian, Frantz,
  Kantarcioglu, and Yao}{Rahaman et~al\mbox{.}}{2019}]%
        {rahaman2019cryptoguard}
\bibfield{author}{\bibinfo{person}{Sazzadur Rahaman}, \bibinfo{person}{Ya
  Xiao}, \bibinfo{person}{Sharmin Afrose}, \bibinfo{person}{Fahad Shaon},
  \bibinfo{person}{Ke Tian}, \bibinfo{person}{Miles Frantz},
  \bibinfo{person}{Murat Kantarcioglu}, {and} \bibinfo{person}{Danfeng Yao}.}
  \bibinfo{year}{2019}\natexlab{}.
\newblock \showarticletitle{Cryptoguard: High precision detection of
  cryptographic vulnerabilities in massive-sized java projects}. In
  \bibinfo{booktitle}{\emph{ACM SIGSAC Conference on Computer and
  Communications Security}}. \bibinfo{pages}{2455--2472}.
\newblock


\bibitem[\protect\citeauthoryear{Zhang, Chen, Diao, Guo, Weng, and Zhang}{Zhang
  et~al\mbox{.}}{2019}]%
        {zhang2019cryptorex}
\bibfield{author}{\bibinfo{person}{Li Zhang}, \bibinfo{person}{Jiongyi Chen},
  \bibinfo{person}{Wenrui Diao}, \bibinfo{person}{Shanqing Guo},
  \bibinfo{person}{Jian Weng}, {and} \bibinfo{person}{Kehuan Zhang}.}
  \bibinfo{year}{2019}\natexlab{}.
\newblock \showarticletitle{CryptoREX: Large-scale Analysis of Cryptographic
  Misuse in IoT Devices}. In \bibinfo{booktitle}{\emph{International Symposium
  on Research in Attacks, Intrusions and Defenses (RAID)}}.
  \bibinfo{publisher}{{USENIX} Association}, \bibinfo{pages}{151--164}.
\newblock
\showISBNx{978-1-939133-07-6}


\end{thebibliography}
\balance

\end{document}